\documentclass{aip-cp}

\usepackage[numbers]{natbib}
\usepackage{rotating}
\usepackage{graphicx}
\usepackage{subcaption}

\usepackage{lineno}

\newcommand{\seboss}{E-BOSS catalogue release 2}  
\newcommand{\hess}{H.E.S.S.}

\newcommand{\zetaoph}{$\zeta$~Ophiuchi }
\newcommand{\bd}{BD$+43^\circ3654$}

\begin{document}

\title{Observations of Bow Shocks of Runaway Stars with H.E.S.S.}

\author[aff1]{A. Schulz\corref{cor1}}

\author[aff1]{M. Haupt}
\author[aff1]{S. Klepser}
\author[aff1]{S. Ohm }
\author{for the H.E.S.S. Collaboration}

\affil[aff1]{DESY, Platanenalee 6, D-15738 Zeuthen, Germany}
\corresp[cor1]{Corresponding author: anneli.schulz@desy.de}

\maketitle

\begin{abstract}
Runaway stars form bow shocks by sweeping up interstellar matter in their direction of motion. Theoretical models predict a spectrally wide non-thermal component reaching up to gamma-ray energies at a flux level detectable with current instruments. They were motivated by a detection of non-thermal radio emission from the bow shock of BD$+43^\circ3654$ and a possible detection of non-thermal X-rays from AE Aurigae. A search in the high-energy regime using data from \textit{Fermi}-LAT resulted in flux upper limits for 27 candidates listed in the first E-BOSS catalogue. We perform the first systematic search for TeV emission from bow shocks of runaway stars. Using all available archival H.E.S.S. I data we search for very-high-energy gamma-ray emission at the positions of bow shock candidates listed in the second E-BOSS catalogue. This catalogue comprises 73 bow shock candidates, 32 of which have been observed with the H.E.S.S. telescopes. None of the observed bow shock candidates shows significant emission in the H.E.S.S. energy range. The resulting upper limits are used to constrain current models for non-thermal emission from these objects.

\end{abstract}	

\section{INTRODUCTION}
\begin{figure}[h!]
  \centering
	\includegraphics[width=0.7\textwidth]{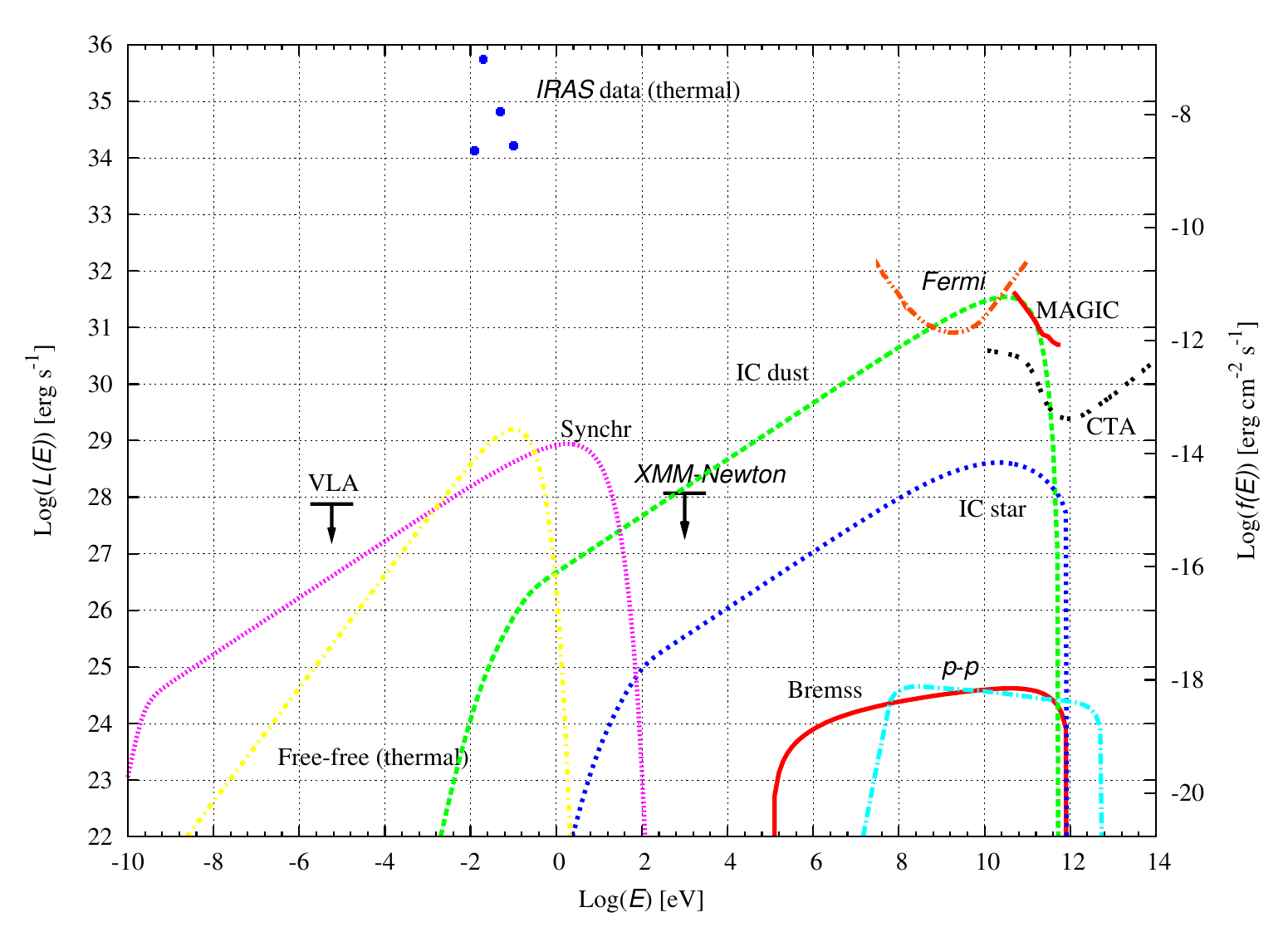} 
  \caption{Prediction for thermal and non-thermal emission from \zetaoph from \citet{Valle_model_zeta}. }%
  \label{fig:zetaoph}%
\end{figure}

Fast moving stars, so-called runaway stars, can create bow shocks in their direction of motion when passing through the interstellar medium with supersonic velocities. These objects are typically detected at infrared wavelengths. The emission in this energy range originates from thermal emission produced by the heated ISM dust. The latest catalogue of stellar bow shocks based on infrared data, the Extensive stellar BOw Shock Survey (E-BOSS release 2; \citet{eboss_r2}), lists 73 candidates. \\ 
The first and so far only detection of non-thermal emission was published by \citet{benaglia_bd} who detected non-thermal radio emission from \bd. They also presented an emission model predicting non-thermal X-ray and gamma-ray emission at a level detectable by current instruments. This model was developed further and applied to other candidates, for example \zetaoph being a prominent candidate due to its proximity. The model prediction for \zetaoph by \citet{Valle_model_zeta} is shown in Figure\,\ref{fig:zetaoph}. \\
Several attempts to observe non-thermal emission followed: radio and X-ray observations of \zetaoph (\citet{vla_skysurvey} and \citet{hasinger_xmm_ul_2001}, respectively), X-ray observations of \bd~by \citet{terada_bd}, and X-ray observations of AE Aurigae by \citet{ae_aurigae_LopezSantiago} and \citet{pereira_modelling}. All resulted in non-detections and subsequently upper limits on the emission. \\
A systematic study on high-energy gamma-ray emission with \textit{Fermi}-LAT was performed by \citet{Schulz_fermi} on 27 bow shock candidates, resulting in upper limits, too. In the case of \zetaoph these limits are a factor $\sim5$ below the model predictions. \\

\section{Bow Shock Sample}
\begin{figure}%
  \centering
	\begin{tabular}{ p{0.9\textwidth} }
		\includegraphics[width=0.9\textwidth]{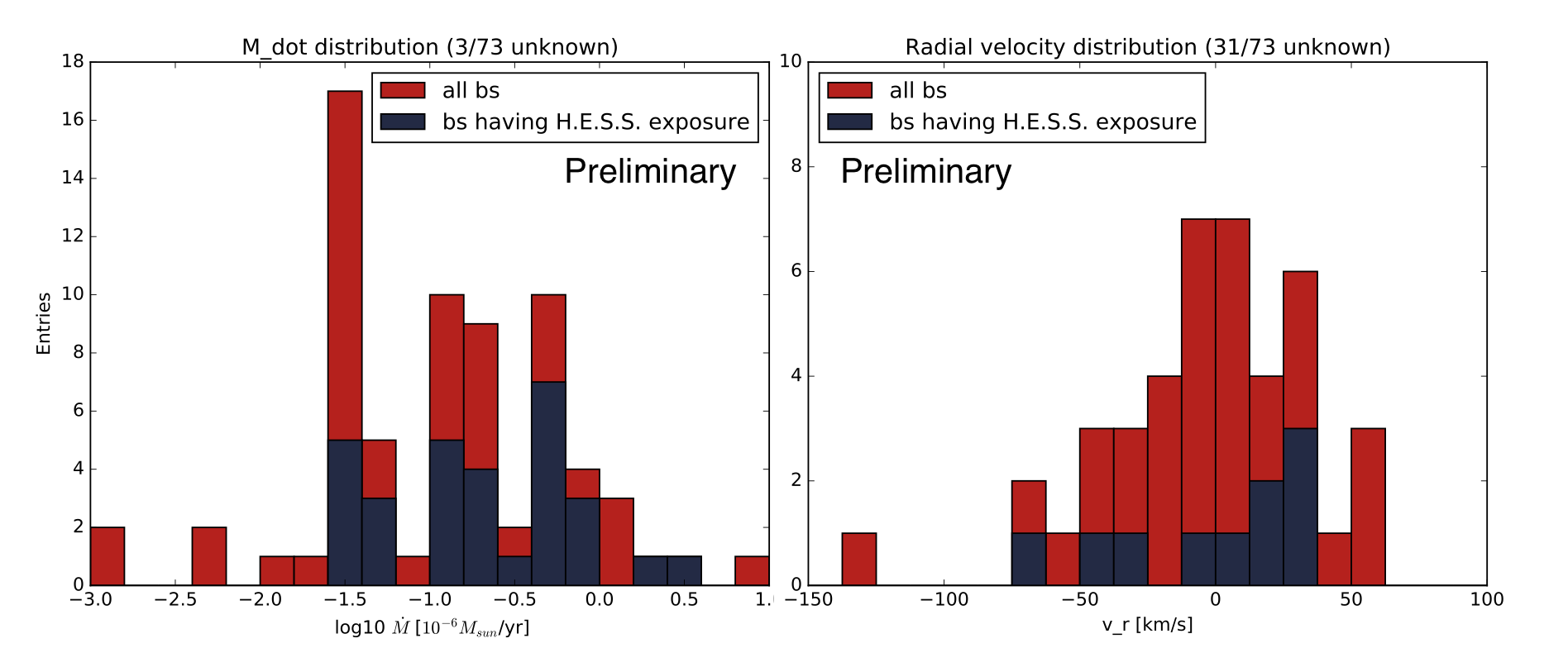} 	\\
		\includegraphics[width=0.9\textwidth]{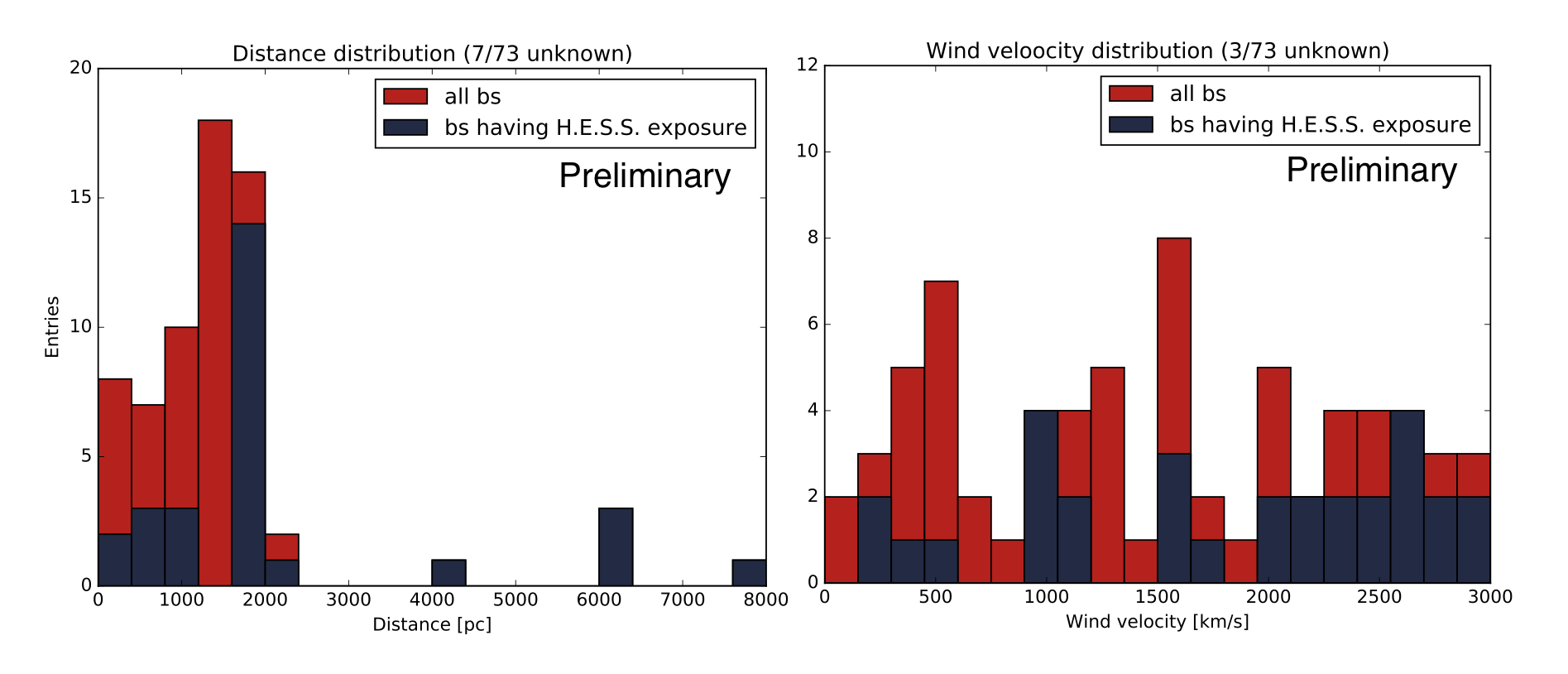}	
		\end{tabular}		
  \caption{Distribution of bow shock parameters, top row: the mass-loss rate $\dot{M}$ and radial velocity $ v_\mathrm{rad} $; bottom row: distance $d$ and wind velocity $ v_\mathrm{wind} $ for the \seboss~sample (in red) and the bow shocks (bs) with \hess~exposure (in blue). Data from \citet{eboss_r2}.  }%
  \label{fig:distris}%
\end{figure}

The basis of this population study is the \seboss, \cite{eboss_r2}. 32 of the 73 candidates of this catalogue have been observed with the \hess~telescopes. The most important parameters are the mass-loss rate $\dot{M}$, radial velocity $ v_\mathrm{rad} $, distance $d$ and wind velocity $ v_\mathrm{wind} $, the distributions in Figure\,\ref{fig:distris} show the diversity of the sample. 

\section{\hess~Analyses}
\begin{figure}[h!]
  \centering
	\begin{tabular}{ p{0.5\textwidth} p{0.5\textwidth} }
		\includegraphics[width=0.5\textwidth]{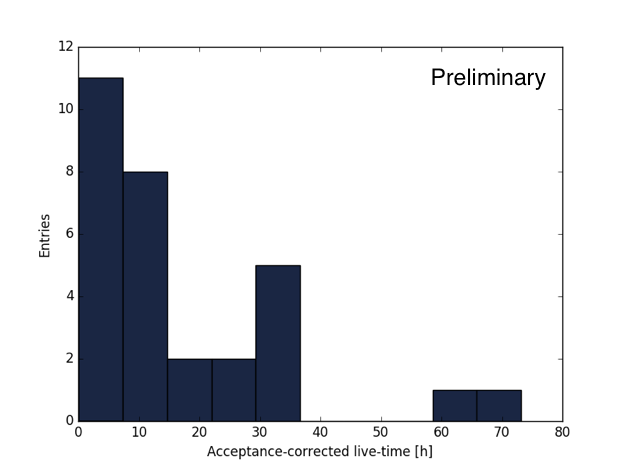} 
			\put(-200,4){a)}
	&
		\includegraphics[width=0.5\textwidth]{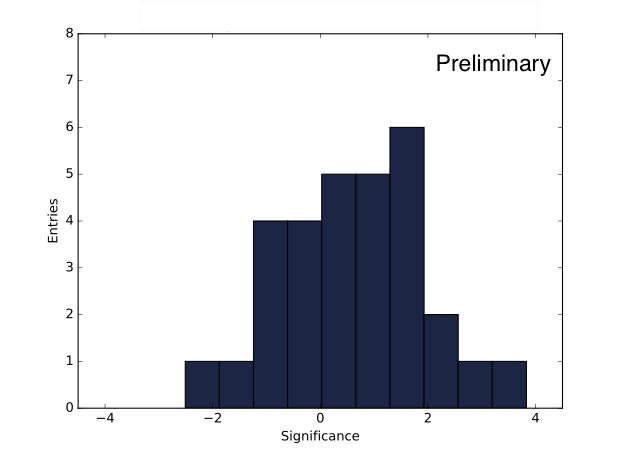}		
			\put(-200,4){b)}

	\end{tabular}
  \caption{Live-time (a) and significance (b) distribution of the \hess~bow shock sample.  }%
  \label{fig:livetime_sig}%
\end{figure}

\begin{figure}[h!]
  \centering
	\includegraphics[width=0.9\textwidth]{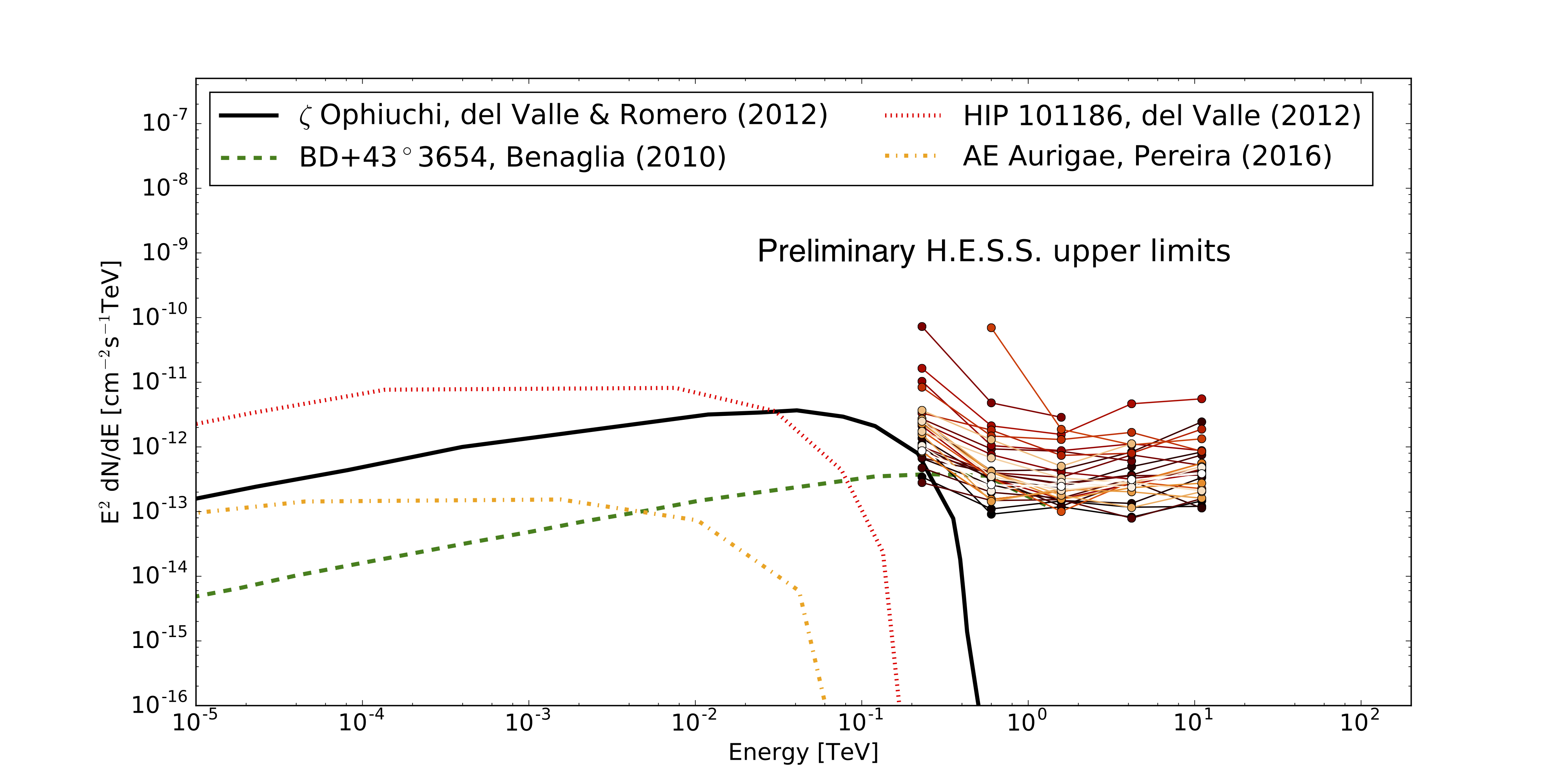} 
  \caption{Upper limits for 32 bow shocks presented in this work together with model predictions for four bow shocks, none of which is in the \hess~bow shock sample.  }%
  \label{fig:ul}%
\end{figure}

The status of the \hess~Cherenkov telescope system was presented in this conference \cite{chaves_hess}. It is clear that the imaging atmospheric Cherenkov telescopes are leaving the time of single source discoveries and entering the era of population studies. This is reflected in this contribution as well as the \hess~galactic plane survey presented in this conference \cite{donath_hgps}, the population study on pulsar wind nebulae presented in this conference  \cite{klepser_pwn} and more detailed in \citet{pwnpop}, and the supernova remnant population study (\citet{snrpop}).\\
32 bow shock candidates have been observed with \hess, the acceptance-corrected live-time distribution is shown in Figure\,\ref{fig:livetime_sig}  a). All observed bow shock candidates were analysed after a cut on standard quality selection parameters with the Impact analysis method described in \citet{parsons_impact_2014}. The cross-check analyses with the methods described in \citet{2009APh....32..231D} yielded compatible results. \\
None of the bow shocks is detected significantly, the distribution of significances is shown in Figure\,\ref{fig:livetime_sig}  b). Consequently flux upper limits are calculated in five energy bins using an index of 2.5 and a confidence level of 99\%. These are shown in Figure\,\ref{fig:ul} together with model predictions for four bow shocks, all not part of the \hess~sample. 

The upper limits can be used to constrain the fraction of the kinetic wind power which is converted into the production of VHE gamma rays, details on the calculation are outlined in  \citet{hess_bs_paper}.

\section{CONCLUSION}
We presented a study on the VHE emission of bow shocks created by runaway stars. 32 bow shock candidates have been observed with \hess, constituting more than 40\% of the candidates in the second release of the E-BOSS catalogue. No significant emission is detected from any of the source candidates, consequently upper limits on the flux are calculated. These non-detections together with the non-detections in \citet{Schulz_fermi} question the level of gamma-ray emission from these objects. The publication including final quantitative statements is forthcoming \citet{hess_bs_paper}. The future Cherenkov Telescope Array (CTA, \citet{CTAHinton2013}), with its improved sensitivity and resolution, will give more insights in the nature of bow shocks of runaway stars.

\section{ACKNOWLEDGMENTS}
The support of the Namibian authorities and of the University of Namibia in facilitating the construction and operation of H.E.S.S. is gratefully acknowledged, as is the support by the German Ministry for Education and Research (BMBF), the Max Planck Society, the German Research Foundation (DFG), the French Ministry for Research, the CNRS-IN2P3 and the Astroparticle Interdisciplinary Programme of the CNRS, the U.K. Science and Technology Facilities Council (STFC), the IPNP of the Charles University, the Czech Science Foundation, the Polish Ministry of Science and Higher Education, the South African Department of Science and Technology and National Research Foundation, the University of Namibia, the Innsbruck University, the Austrian Science Fund (FWF), and the Austrian Federal Ministry for Science, Research and Economy, and by the University of Adelaide and the Australian Research Council. We appreciate the excellent work of the technical support staff in Berlin, Durham, Hamburg, Heidelberg, Palaiseau, Paris, Saclay, and in Namibia in the construction and operation of the equipment. This work benefited from services provided by the H.E.S.S. Virtual Organisation, supported by the national resource providers of the EGI Federation.\\
This publication makes use of data products from the Wide-field Infrared Survey Explorer, which is a joint project of the University of California, Los Angeles, and the Jet Propulsion Laboratory/California Institute of Technology, funded by the National Aeronautics and Space Administration.


\bibliographystyle{aipnum-cp}%
\bibliography{references_bowshocks}%

\end{document}